\documentclass[12pt,thmsa]{article}

\input{tcilatex}

\begin{document}

\begin{center}
\textbf{Unidirectional Propagation of an Ultra-Short Electromagnetic Pulse\\[%
0pt]
in a Resonant Medium with High Frequency Stark Shift}

\bigskip

\vspace{0.5in}Jean-Guy Caputo$^{a}$\label{a}\footnote{%
electronic address: caputo@insa-rouen.fr, corresponding author} and Andrei
I. Maimistov$^{b}$\footnote{%
electronic address: maimistov@pico.mephi.ru}

\vspace{0.2cm} $^{a}$Laboratoire de Math\'ematiques, INSA de Rouen, B.P. 8,
76131 Mont-Saint-Aignan cedex, France, \vspace{0.2cm}

$^{b}$Department of Solid State Physics, Moscow Engineering Physics
Institute, Kashirskoe sh. 31, Moscow, 115409 Russia,
\end{center}

\bigskip

\begin{center}
\textbf{ABSTRACT}
\end{center}

\bigskip

We consider in the unidirectional approximation the propagation of an ultra
short electromagnetic pulse in a resonant medium consisting of molecules
characterized by a transition operator with both diagonal and non-diagonal
matrix elements. We find the zero-curvature representation of the reduced
Maxwell-Bloch equations in the sharp line limit. This can be used to develop
the inverse scattering transform method to solve these equations. Finally we
obtain two types of exact traveling pulse solutions, one with the usual
exponential decay and another with an algebraic decay.

\textit{PACS}: 42.65. Tg

\textit{Keywords}: ultra-short pulses, two-level atoms model, high frequency
Stark effect, inverse scattering transform method, steady state pulse,
soliton.

\section{Introduction}

If the duration of an electromagnetic pulse is less than all the relaxation
times in a medium, then the propagation of such a pulse is accompanied only
by stimulated absorption and re-emission. These electromagnetic pulses we
will refer to as ultra-short (USP). Usually the study of the propagation of
USP within the framework of two-level atoms \cite{R1,R2} assumes that the
diagonal matrix elements of the dipole moment operator are zero (see also
the reviews \cite{R3,R4,R5}). This model could also describe the very short
electromagnetic pulses containing a few optical cycles up to a half cycle 
\cite{R6,R7,R8,R9}.

An important approximation for the electromagnetic field called \emph{the
slowly varying envelope approximation}, assumes that the pulse envelope is a
function slowly varying in space and time in comparison to the carrier
monochromatic wave. In this approximation the Bloch equations describing the
evolution of two-level atoms and the equation for the pulse envelope can be
solved by t\emph{he inverse scattering transform} (IST) method \cite
{R10,R11,R12}. There are soliton and multi-solitons solutions, which
describe the propagation and (elastic) interaction between USP in a resonant
medium.

Second in importance is\emph{\ the unidirectional wave approximation}. This
corresponds to the propagation of the electromagnetic wave only in one of
the two possible directions. The Maxwell and Bloch equations in this
approximation allow to consider the USP propagation without imposing
limitations on the pulse duration. As in the first case, the reduced
Maxwell-Bloch equations can be solved via the IST method. Here the soliton
corresponds to an extremely short pulse of the electromagnetic field, which
contains half an optical cycle. The breather solutions correspond to pulses
containing a few optical cycles. As already noted, two-level atoms media
have been considered with zero diagonal matrix elements of the dipole moment
operator. In a medium with a linear Stark shift, this assumption should be
revised. The constant diagonal matrix elements of the dipole moment operator
can be also induced by the external constant electric field. In an
electromagnetic wave the resonant levels of this medium will be shifted. In
analogy to a Kerr medium, for which a high frequency electromagnetic field
results into a high frequency Kerr effect, a medium possessing a high
frequency Stark effect may be referred to as a \emph{Stark medium}. In Ref. 
\cite{R13} the interaction of a few-cycle electromagnetic pulses with a
two-level Stark medium was considered in detail. Unlike \cite{R13} here we
consider a simpler model, where we assume that all the matrix elements of
the dipole moment operator are parallel to the linearly polarized electric
field vector, neglect relaxation processes and the inhomogeneous broadening
of the resonant lines, and assume unidirectional waves. This specific model
differs from the standard two-level atoms model in a minimal degree. This
results in a new kind of steady state propagation of the USP of extremely
short duration. Furthermore, this model admits a zero-curvature
representation, which can be used as a base to obtain exact solution by the
IST method. After describing the reduced Maxwell-Bloch equations in section
2, we give a zero-curvature representation in section 3 and find steady
state pulse solutions in section 4.

\section{The model and the principal equations}

Following \cite{R11}\ we consider a plane electromagnetic wave propagating
into a resonant medium consisting of molecules characterized by the operator
of the dipole transition between resonant energy levels and let this
operator have both non-diagonal and diagonal matrix elements. In the
two-level approximation the Hamiltonian of the considered model can be
written as 
\[
\hat{H}=\frac{\hbar \omega _{0}}{2}\left( 
\begin{array}{cc}
-1 & 0 \\ 
0 & 1
\end{array}
\right) -\left( 
\begin{array}{cc}
d_{11}E & d_{12}E \\ 
d_{21}E & d_{22}E
\end{array}
\right) , 
\]
where $E$ is the strength of electric field of the electromagnetic wave. The
polarization of the medium is $P=n_{A}p$, where $n_{A}$ is the density of
the molecules and the polarizability $p$ is given by the expression $p=$tr$%
\hat{\rho }\hat{d}=\rho _{11}d_{11}+\rho _{22}d_{22}+\rho _{12}d_{21}+\rho
_{21}d_{12}$. We will consider the propagation of short electromagnetic
pulses, for which all relaxation processes can be neglected. Hence, the
density matrix $\hat{\rho }$ obeys the constraint $\rho _{11}+\rho _{22}=1$.
Taking this relation into account, the polarizability can be written as 
\[
p=\frac{1}{2}\left( d_{11}+d_{22}\right) +\frac{1}{2}\left(
d_{11}-d_{22}\right) (\rho _{11}-\rho _{22})+\rho _{12}d_{21}+\rho
_{21}d_{12}. 
\]
The evolution of the elements of $\hat{\rho }$\ is given by the equation $%
i\hbar \partial \hat{\rho }/\partial t=\hat{H}\hat{\rho }-\hat{\rho }\hat{H}$%
\ and yields the Bloch equations.

In a scalar form the Maxwell equations lead to 
\begin{equation}
\frac{\partial ^{2}E}{\partial z^{2}}-\frac{1}{c^{2}}\frac{\partial ^{2}E}{%
\partial t^{2}}=\frac{4\pi n_{A}}{c^{2}}\frac{\partial ^{2}}{\partial t^{2}}%
\left\langle p\right\rangle ,  \label{eq21}
\end{equation}
where the angular brackets stand for averaging over all molecules.

It is convenient to introduce the Bloch vector by defining its components 
\[
r_{1}=\rho _{12}+\rho _{21},\quad r_{2}=-i(\rho _{12}-\rho _{21}),\quad
r_{3}=\rho _{22}-\rho _{11}. 
\]
On can choose a constant phase for the matrix elements of the density matrix 
$\hat{\rho}$\ and the dipole operator $\hat{d}$ so that $d_{12}=d_{21}$. The
polarizability $p$ of the resonant medium can be written in terms of the
Bloch vector components as 
\begin{equation}
p=\frac{1}{2}\left( d_{11}+d_{22}\right) +\frac{1}{2}\left(
d_{22}-d_{11}\right) r_{3}+d_{12}r_{1}.  \label{eq22}
\end{equation}
In this expression the first term corresponds to the constant polarizability
of the molecules. As it gives no contribution to the radiation, it may be
omitted. Furthermore, after averaging over all atoms it must be $%
\left\langle d_{11}+d_{22}\right\rangle =0$ if one assumes that there is no
constant polarization of the medium in the absence of an electromagnetic
field.

So the total system of equations describing the model under consideration
can be written in the following form 
\[
\frac{\partial ^{2}E}{\partial z^{2}}-\frac{1}{c^{2}}\frac{\partial ^{2}E}{%
\partial t^{2}}=\frac{4\pi n_{A}}{c^{2}}\frac{\partial ^{2}}{\partial t^{2}}%
\left\langle \frac{1}{2}\left( d_{22}-d_{11}\right)
r_{3}+d_{12}r_{1}\right\rangle , 
\]
\[
\frac{\partial r_{1}}{\partial t}=-\left[ \omega _{0}+\left(
d_{11}-d_{22}\right) E/\hbar \right] r_{2}, 
\]
\[
\frac{\partial r_{2}}{\partial t}=-\left[ \omega _{0}+\left(
d_{11}-d_{22}\right) E/\hbar \right] r_{1}+2(dE/\hbar )r_{3}, 
\]
\[
\frac{\partial r_{3}}{\partial t}=-2(dE/\hbar )r_{2}. 
\]
This system differs from the well-known Maxwell-Bloch equations \cite{R2}- 
\cite{R5} by terms containing the parameter $\left( d_{11}-d_{22}\right) $.
If $n_{A}\left| d_{11}-d_{22}\right| $\ and $n_{A}\left| d_{12}\right| $ are
small, then one can neglect the reflected wave and consider the
unidirectional propagation of the electromagnetic wave \cite{R4,R9}. The
system of the Maxwell-Bloch equations is then reduced to 
\begin{equation}
\frac{\partial E}{\partial z}+\frac{1}{c}\frac{\partial E}{\partial t}=-%
\frac{2\pi n_{A}}{c^{2}}\frac{\partial }{\partial t}\left\langle \frac{1}{2}%
\left( d_{22}-d_{11}\right) r_{3}+d_{12}r_{1}\right\rangle ,  \label{eq23}
\end{equation}
\begin{equation}
\frac{\partial r_{1}}{\partial t}=-\left[ \omega _{0}+\left(
d_{11}-d_{22}\right) E/\hbar \right] r_{2},  \label{eq24}
\end{equation}
\begin{equation}
\frac{\partial r_{2}}{\partial t}=-\left[ \omega _{0}+\left(
d_{11}-d_{22}\right) E/\hbar \right] r_{1}+2(dE/\hbar )r_{3},  \label{eq25}
\end{equation}
\begin{equation}
\frac{\partial r_{3}}{\partial t}=-2(dE/\hbar )r_{2}.  \label{eq26}
\end{equation}
Let us consider here only the sharp line limit of the model. Introduce new
dimensionless variables and field 
\[
\tau =\omega _{0}(t-z/c),\quad \zeta =z/L_{ab},\quad q=2dE/\hbar \omega
_{0}, 
\]
where $L_{ab}^{-1}=4\pi n_{A}d_{12}^{2}(\hbar c)^{-1}$ and the parameter $%
\mu =\left( d_{11}-d_{22}\right) /2d_{12}$. The reduced Maxwell- Bloch
equations (2.3)-(2.6) take the form: 
\begin{equation}
r_{1,\tau }=-(1+\mu q)r_{2},\quad r_{2,\tau }=(1+\mu q)r_{1}+qr_{3},\quad
r_{3,\tau }=-qr_{2},  \label{eq27}
\end{equation}
\begin{equation}
q_{,\zeta }=-(r_{1}-\mu r_{3})_{,\tau },  \label{eq28}
\end{equation}
The last equation can be rewritten as 
\begin{equation}
q_{,\zeta }=r_{2}.  \label{eq29}
\end{equation}
We will assume the general initial conditions where the medium is initially
at rest so that $q=0,$ $r_{1}=r_{2}=0,$ $r_{3}=-1$, at $\tau \longrightarrow
-\infty $.

From the Bloch equations (\ref{eq27}) we obtain $\left(
r_{1}^{2}+r_{2}^{2}+r_{3}^{2}\right) _{,\tau }=0.$\ and using the initial
conditions we get the following integral of motion 
\begin{equation}
r_{1}^{2}+r_{2}^{2}+r_{3}^{2}=1.  \label{eq210}
\end{equation}
We now proceed to give the zero-curvature representation of the system (\ref
{eq27})-(\ref{eq29}). We will see that the $\mu q$\ term which makes the
difference from the classical reduced Maxwell-Bloch equations \cite
{R4,R5,R14} will generate complications of traditional IST method. We will
also find two types of traveling pulse solutions which differ from the ones
of the two classical reduced Maxwell-Bloch systems.

\bigskip

\section{The zero-curvature representation}

The inverse scattering transform method is based on considering the pair of
linear equations \cite{R15}: 
\begin{equation}
\psi _{,\tau }=\hat{U}\psi ,\qquad \psi _{,\zeta }=\hat{V}\psi ,
\label{eq31}
\end{equation}
where 
\[
\hat{V}=\left( 
\begin{array}{cc}
A & B \\ 
C & -A
\end{array}
\right) ,\qquad \hat{U}=\left( 
\begin{array}{cc}
U_{11} & U_{12} \\ 
U_{21} & U_{22}
\end{array}
\right) . 
\]
The compatibility condition for these equations is $\hat{U}_{\zeta }=\hat{V}%
_{,\tau }+\hat{V}\hat{U}-\hat{U}\hat{V}$, which yields the following system
of equations 
\begin{equation}
A_{,\tau }+U_{21}B-U_{12}C=U_{11,\zeta },  \label{eq32}
\end{equation}
\begin{equation}
B_{,\tau }+\left( U_{22}-U_{11}\right) B+2U_{12}A=U_{12,\zeta },
\label{eq33}
\end{equation}
\begin{equation}
C_{,\tau }-\left( U_{22}-U_{11}\right) C-2U_{21}A=U_{21,\zeta }.
\label{eq34}
\end{equation}
It should be pointed that the reduced Maxwell-Bloch equations (\ref{eq27})-(%
\ref{eq29}) can be represented in another form: 
\begin{equation}
i\psi _{1,\tau }=-(1/2)(1+\mu q)\psi _{1}-(1/2)q\psi _{2},
\end{equation}
\begin{equation}
i\psi _{2,\tau }=-(1/2)q\psi _{1}+(1/2)(1-\mu q)\psi _{1},
\end{equation}
\begin{equation}
q_{,\zeta }=i\left( \psi _{1}^{\ast }\psi _{2}-\psi _{1}\psi _{2}^{\ast
}\right) ,
\end{equation}
where the components of the Bloch vector are constructed from the functions $%
\psi $\ according to the rule: 
\[
r_{1}=\left( \psi _{1}^{\ast }\psi _{2}+\psi _{1}\psi _{2}^{\ast }\right)
,\quad r_{2}=i\left( \psi _{1}^{\ast }\psi _{2}-\psi _{1}\psi _{2}^{\ast
}\right) ,\quad r_{3}=\left| \psi _{2}\right| ^{2}-\left| \psi _{1}\right|
^{2}. 
\]
This suggests that the U-matrix of the zero-curvature representation for the
reduced Maxwell- Bloch equations (\ref{eq27})-(\ref{eq29}) could be written
as 
\begin{equation}
\hat{U}=\left( 
\begin{array}{cc}
-i\lambda +f_{1}(\lambda )q & g_{1}(\lambda )q \\ 
g_{1}(\lambda )q & i\lambda -f_{2}(\lambda )q
\end{array}
\right) ,
\end{equation}
where $f_{1,2}(\lambda )$\ and $g_{1,2}(\lambda )$\ are unknown functions of
the spectral parameter $\lambda $. Taking into account this U-matrix, the
compatibility conditions (\ref{eq32})-(\ref{eq34}) can be written as 
\begin{equation}
A_{,\tau }+g_{2}(\lambda )qB-g_{1}(\lambda )qC=f_{1}(\lambda )q_{,\zeta },
\label{eq39}
\end{equation}
\begin{equation}
B_{,\tau }+\left( 2\eta -f(\lambda )q\right) B+2g_{1}(\lambda
)qA=g_{1}(\lambda )q_{,\zeta },  \label{eq310}
\end{equation}
\begin{equation}
C_{,\tau }-\left( 2\eta -f(\lambda )q\right) C-2g_{2}(\lambda
)qA=g_{2}(\lambda )q_{,\zeta },  \label{eq311}
\end{equation}
where $f(\lambda )=f_{1}(\lambda )+f_{2}(\lambda )$, and $\eta =i\lambda $.
The system of equations (\ref{eq39})-(\ref{eq311}) must coincide with the
reduced Maxwell-Bloch equations (\ref{eq27})-(\ref{eq29}).

It should be noted that the system of equations (\ref{eq27})-(\ref{eq29})
contains the partial derivatives only of the Bloch vector components with
respect to $\tau $. Hence one can find $B$\ and $C$ from (\ref{eq310}) and (%
\ref{eq311}) using the following expansions 
\[
B=b_{1}r_{1}+b_{2}r_{2}+b_{3}r_{3},\quad C=c_{1}r_{1}+c_{2}r_{2}+c_{3}r_{3}, 
\]
where $b_{1,2,3}$ and $c_{1,2,3}$ are unknown functions of $\lambda $.
Substituting these expansions into equations (\ref{eq310}) and (\ref{eq311}%
), and equating the coefficients of $r_{1},r_{2}$, $r_{3}$\ and\ $q$\ in the
resulting equations, we obtain the following relations between the $%
b_{1,2,3} $ and $c_{1,2,3}$%
\begin{equation}
b_{2}+2\eta b_{1}=0,\quad -b_{1}+2\eta b_{2}=g_{1},\quad b_{3}=0,
\label{eq312}
\end{equation}
\begin{equation}
c_{2}-2\eta c_{1}=0,\quad -c_{1}-2\eta c=g_{2},\quad c_{3}=0,  \label{eq313}
\end{equation}
\begin{equation}
2g_{1}A+b_{1}\left( -\mu r_{2}-fr_{1}\right) +b_{2}\left( \mu
r_{1}+r_{3}-fr_{2}\right) =0,  \label{eq314}
\end{equation}
\begin{equation}
-2g_{2}A+c_{1}\left( -\mu r_{2}+fr_{1}\right) +c_{2}\left( \mu
r_{1}+r_{3}+fr_{2}\right) =0.  \label{eq315}
\end{equation}
From (\ref{eq312}) and (\ref{eq313}) it follows that 
\[
b_{1}=\frac{-g_{1}}{1+4\eta ^{2}},\quad b_{2}=\frac{2\eta g_{1}}{1+4\eta ^{2}%
},\quad c_{1}=\frac{-g_{2}}{1+4\eta ^{2}},\quad c_{2}=\frac{-2\eta g_{2}}{%
1+4\eta ^{2}}. 
\]
Substitution of these expressions into (\ref{eq314}) and (\ref{eq315}) leads
to two equations for the matrix element $A$. From (3.14) one gets 
\[
2A+\left( 1+4\eta ^{2}\right) ^{-1}\left[ 2\eta r_{3}+\left( 2\eta \mu
+f\right) r_{1}+\left( \mu -2\eta f\right) r_{2}\right] =0, 
\]
and from (3.15) one gets 
\[
2A+\left( 1+4\eta ^{2}\right) ^{-1}\left[ 2\eta r_{3}+\left( 2\eta \mu
+f\right) r_{1}-\left( \mu -2\eta f\right) r_{2}\right] =0. 
\]
Both expressions agree if $\mu -2\eta f=0$. Thus the auxiliary function $%
f(\lambda )$\ is specified: $f(\lambda )=\mu /2\eta $, and the expressions
for $B$, $C$ and $A$ are determined: 
\begin{equation}
B=\frac{g_{1}}{1+4\eta ^{2}}\left( -r_{1}+2\eta r_{2}\right) ,\quad C=\frac{%
g_{2}}{1+4\eta ^{2}}\left( -r_{1}-2\eta r_{2}\right) ,  \label{eq316}
\end{equation}
\begin{equation}
A=\frac{-1}{1+4\eta ^{2}}\left[ \eta r_{3}+\mu \left( \eta +\frac{1}{4\eta }%
\right) r_{1}\right] .  \label{eq317}
\end{equation}

To determine the auxiliary function $g_{1,2}(\lambda )$, equation (3.9) has
to be used. We assume the equality $f_{1}(\lambda )=f_{2}(\lambda )$ so that 
$f_{1}(\lambda )=\mu /4\eta $. Substitution of the equations (\ref{eq316})
and (\ref{eq317}) into (\ref{eq39}) leads to the following relation 
\[
\eta r_{3,\tau }+\mu \left( \eta +\frac{1}{4\eta }\right) r_{1,\tau }+\frac{%
\mu }{4\eta }\left( 1+4\eta ^{2}\right) q_{,\zeta }=g_{1}g_{2}4\eta qr_{2}. 
\]
Using equations (\ref{eq27}) and (\ref{eq29}) this becomes 
\begin{equation}
\eta +\frac{\mu ^{2}}{4\eta }\left( 1+4\eta ^{2}\right) +4\eta g_{1}g_{2}=0.
\label{eq318}
\end{equation}
Now, we can require any correlation between $g_{1}$\ and $g_{2}$, for
instance, $g_{1}(\lambda )=-g_{2}(\lambda )=g(\lambda )$. From (\ref{eq318})
it then follows immediately that 
\begin{equation}
g(\lambda )=\frac{1}{2}\left[ 1+\frac{\mu ^{2}}{4\eta ^{2}}\left( 1+4\eta
^{2}\right) \right] ^{1/2}.  \label{ea319}
\end{equation}
Some times it may be more convenient to take $g_{1}(\lambda )=g_{2}(\lambda
)=ig(\lambda )$.

Thus, we have the U-V-matrices for a zero-curvature representation of the
reduced Maxwell- Bloch equations (\ref{eq27}) and (\ref{eq29}) describing
the USP propagation in Stark medium assuming a model of two-level atoms.
They are \smallskip 
\begin{equation}
\hat{U}=\left( 
\begin{array}{cc}
-\eta +\left( \mu /4\eta \right) q & \frac{1}{2}\left[ 1+\left( \mu /2\eta
\right) ^{2}\left( 1+4\eta ^{2}\right) \right] ^{1/2}q \\ 
-\frac{1}{2}\left[ 1+\left( \mu /2\eta \right) ^{2}\left( 1+4\eta
^{2}\right) \right] ^{1/2}q & \eta -\left( \mu /4\eta \right) q
\end{array}
\right) ,  \label{eq320}
\end{equation}
\smallskip 
\[
\hat{V}=\frac{1}{2(1+4\eta ^{2})}\times 
\]
\begin{equation}
\left( 
\begin{array}{cc}
-2\left[ \eta r_{3}+\mu \left( \eta +\left( 1/4\eta \right) \right) r_{1}%
\right] & \left[ 1+\left( \mu /2\eta \right) ^{2}\left( 1+4\eta ^{2}\right) 
\right] ^{1/2}\left( 2\eta r_{2}-r_{1}\right) \\ 
\left[ 1+\left( \mu /2\eta \right) ^{2}\left( 1+4\eta ^{2}\right) \right]
^{1/2}\left( 2\eta r_{2}+r_{1}\right) & 2\left[ \eta r_{3}+\mu \left( \eta
+\left( \mu /4\eta \right) \right) r_{1}\right]
\end{array}
\right) ,  \label{eq321}
\end{equation}
where $\eta =i\lambda $. It should be noted that if $\mu \longrightarrow 0$,
then we obtain the well-known U-V representation for the classical reduced
Maxwell- Bloch equations (see \cite{R11,R12}).

\section{Steady state solutions of the Maxwell-Bloch equations}

To obtain the equations describing the propagation of stationary USP one
should assume that the components of the Bloch vector and the normalized
pulse envelope depend on only one variable $\eta \equiv \tau -\beta \zeta
\equiv \omega _{0}(t-z/V)$. Under this assumption the system (\ref{eq27})
and (\ref{eq29}) transforms into the system of ordinary differential
equations 
\begin{equation}
r_{1,\eta }=-(1+\mu q)r_{2},\quad r_{2,\eta }=(1+\mu q)r_{1}+qr_{3},\quad
r_{3,\eta }=-qr_{2},  \label{eq41}
\end{equation}
\begin{equation}
\beta q_{,\eta }=-r_{2}.  \label{eq42}
\end{equation}

Let us consider the solution of equations (\ref{eq41}) and (\ref{eq42})
describing a solitary steady state wave with the boundary conditions $%
q=0,r_{1}=r_{2}=0,r_{3}=-1$\ at $\left| \eta \right| \longrightarrow \infty $
.\newline
From (\ref{eq41}) it follows that 
\begin{equation}
r_{2}=-(r_{1}-\mu r_{3})_{,\eta }  \label{eq43}
\end{equation}
We now take (\ref{eq42}) and the boundary conditions into account to get 
\begin{equation}
r_{1}=\beta q+\mu (1+r_{3}).  \label{eq44}
\end{equation}
From the last equation of (\ref{eq41}) and equation (\ref{eq42}) we obtain 
\begin{equation}
r_{3}=-1+\beta q^{2}/2.
\end{equation}
We define $q_{0}^{2}=4/\beta $\ to renormalize $q$ as $w=q/q_{0}$ and write
all components of the Bloch vector as $r_{1}=4w/q_{0}+2\mu
w,r_{2}=-4w_{,\eta }/q_{0},r_{3}+-1+2w^{2}$. The substitution of the Bloch
vector components into expression (\ref{eq210}) leads at once to the
equation for $w$: 
\begin{equation}
\left( \frac{dw}{d\eta }\right) ^{2}=\left( \frac{q_{0}^{2}}{4}-1\right)
w^{2}-\mu q_{0}w^{3}-\frac{q_{0}^{2}}{4}\left( 1+\mu ^{2}\right) w^{4}.
\label{eq46}
\end{equation}
This equation has a solution in the form of a solitary wave if the factor in
the first term is positive. Let us denote it as $q_{0}^{2}/4-1=\theta ^{-2}$%
\ and introduce the new variable 
\[
y=2\left( q_{0}w\sqrt{1+\mu ^{2}}\right) ^{-1}. 
\]
In terms of this variable equation (\ref{eq46}) takes the form 
\begin{equation}
\left( \frac{dy}{d\eta }\right) ^{2}=\frac{1}{\theta ^{2}}\left( y^{2}-\frac{%
2\mu \theta ^{2}}{\sqrt{1+\mu ^{2}}}y-\theta ^{2}\right) .  \label{eq47}
\end{equation}
Let be $u=y-\mu \theta ^{2}/(1+\mu ^{2})^{1/2}.$. Then equation (\ref{eq47})
can be rewritten as 
\[
\theta \left( \frac{du}{d\eta }\right) =\left( u^{2}-B^{2}\right) ^{1/2}, 
\]
where 
\[
B^{2}=\theta ^{2}\left( 1+\frac{\mu ^{2}\theta ^{2}}{1+\mu ^{2}}\right) . 
\]
The solution of this equation is $u(\eta )=B\cosh \left[ \left( \eta -\eta
_{0}\right) /\theta \right] $, with $\eta _{0}$ as the integrating constant.
Taking this result into account one can obtain that 
\begin{equation}
q\left( \eta \right) =\frac{2}{\theta \left\{ \sqrt{1+\mu ^{2}(1+\theta ^{2})%
}\cosh \left[ \left( \eta -\eta _{0}\right) /\theta \right] +\mu \theta
\right\} }.  \label{eq48}
\end{equation}
This expression represents a one-parameter family of solutions of equations (%
\ref{eq41}) and (\ref{eq42}), with $\theta $ as parameter. It is convenient
to introduce the duration of a stationary USP (\ref{eq48}) $t_{s}$ by the
relation $\theta =t_{s}\omega _{0}$, so that $q\left( \eta \right) $ may be
written as 
\begin{equation}
q\left( \eta \right) =\frac{(2/t_{s}\omega _{0})}{\left\{ \sqrt{1+\mu
^{2}1+(\mu t_{s}\omega _{0})^{2})}\cosh \left[ \left( t-z/V-t_{0}\right)
/t_{s}\right] +\mu t_{s}\omega _{0}\right\} }.  \label{eq49}
\end{equation}
Thus one can obtain the expression for the electric field of this extremely
short steady state pulse: 
\begin{equation}
E_{s}\left( t,z\right) =\frac{(\hbar /t_{s}d)}{\left\{ \sqrt{1+\mu
^{2}1+(\mu t_{s}\omega _{0})^{2})}\cosh \left[ \left( t-z/V-t_{0}\right)
/t_{s}\right] +\mu t_{s}\omega _{0}\right\} }.  \label{eq410}
\end{equation}
From the definition of $\eta $ one can obtain the velocity of propagation $V$
of the USP in the laboratory reference frame 
\begin{equation}
\frac{c}{V}=1+\frac{\left( t_{s}\omega _{0}\right) ^{2}}{1+\left(
t_{s}\omega _{0}\right) ^{2}}\left( \frac{4\pi n_{A}d_{12}^{2}}{\hbar \omega
_{0}}\right) .  \label{eq411}
\end{equation}
This expression for the velocity of the propagation of stationary pulse is
identical to the one found in \cite{R6}. Hence the linear high frequency
Stark effect does not influence the rate of propagation of the steady state
USP.

The expression for the normalized electric field strength (\ref{eq48}) has
been obtained under the condition that $\theta ^{-2}=q_{0}^{2}/4-1>0$. Let
us consider the case $\theta =0$, i.e., $q_{0}=2$. Equation (\ref{eq46})
takes the form 
\begin{equation}
\left( \frac{dw}{d\eta }\right) ^{2}=-2\mu w^{3}-\left( 1+\mu ^{2}\right)
w^{4}.  \label{eq412}
\end{equation}
If $\mu $\ is positive we have only trivial solutions of this equation, $w=0$%
. If $\mu $ is negative there are nontrivial solutions. We rewrite (\ref
{eq412}) as 
\begin{equation}
\left( \frac{dw}{d\eta }\right) ^{2}=w^{3}\left[ 2\left| \mu \right| -\left(
1+\mu ^{2}\right) w\right] .  \label{eq413}
\end{equation}
and introduce $u=2\left| \mu \right| /\left( 1+\mu ^{2}\right) w$ so that
the equation becomes 
\[
\left( \frac{du}{d\eta }\right) ^{2}=\frac{4\mu ^{2}}{\left( 1+\mu
^{2}\right) }\left( u-1\right) . 
\]
The solution of this simple equation is 
\[
u=1+\frac{\mu ^{2}}{\left( 1+\mu ^{2}\right) }\left( \eta -\eta _{0}\right)
^{2}. 
\]
Hence 
\[
y=\frac{1+\mu ^{2}}{2\left| \mu \right| }\left\{ 1+\frac{\mu ^{2}}{1+\mu ^{2}%
}\left( \eta -\eta _{0}\right) ^{2}\right\} , 
\]
and 
\begin{equation}
q_{al}\left( \eta \right) =\frac{4\left| \mu \right| }{\left( 1+\mu
^{2}\right) }\left\{ 1+\frac{\mu ^{2}}{1+\mu ^{2}}\left( \eta -\eta
_{0}\right) ^{2}\right\} ^{-1}.  \label{eq414}
\end{equation}

In terms of physical values, this algebraic steady state USP can be
represented as 
\begin{equation}
E_{al}\left( t,z\right) =\frac{E_{m}}{1+\left[ (t-z/V_{al}-t_{0})/t_{al}%
\right] ^{2}},  \label{eq415}
\end{equation}
where the amplitude of USP $E_{m}$, its duration $t_{al}$, and velocity $%
V_{al}$ are defined as 
\begin{equation}
E_{m}=\frac{4\hbar \omega _{0}(d_{22}-d_{11})}{\sqrt{%
4d_{12}^{2}+(d_{22}-d_{11})^{2}}},\quad t_{al}=\frac{\omega
_{0}(d_{22}-d_{11})}{\sqrt{4d_{12}^{2}+(d_{22}-d_{11})^{2}}},\quad \frac{c}{%
V_{al}}=1+\frac{4\pi n_{A}d_{12}^{2}}{\hbar \omega _{0}}.  \label{eq416}
\end{equation}
The possibility of the propagation of such an algebraic solitary wave is the
distinguishing feature of the Stark medium under consideration here.

\bigskip

\section{Conclusion}

We have introduced and analyzed a model for the propagation of ultra-short
electromagnetic pulses moving in one direction in a Stark medium. This is
described by two-level atoms taking into account the high frequency linear
Stark shift of the energy levels. Two families of exact analytical solutions
of the reduced Maxwell-Bloch equations have been found, which correspond to
steady state pulses propagating with different pulse widths. We find
algebraic solitary waves apart from the usual exponentially decreasing ones.

It was found that the system of reduced Maxwell-Bloch equations admits a
zero-curvature representation in the sharp line limit. Unlike the well-known
Zakharov-Shabat spectral problem the one obtained here shows complicated
analytical properties. Nevertheless we assume that it can be used to develop
the inverse scattering transform method to solve these equations.

There is another class of stationary waves -- cnoidal waves amongst the
solutions of the Maxwell-Bloch equations (\ref{eq23})-(\ref{eq26}), or (\ref
{eq27}) and (\ref{eq29}). These are periodic continuous waves, different
from the solitary waves described above. Since equations (\ref{eq27}) and (%
\ref{eq29}) are valid when the duration of the wave is less or much less
than the relaxation times of the atomic subsystem, cnoidal waves seem to be
mathematical objects lying beyond the physical meaning of the original
equations.

\section{Acknowledgment}

One of the authors (A.I.M.) is grateful to the \textit{Laboratoire de
Mathematiques, INSA de Rouen} for hospitality and support.

\end{document}